\documentstyle[11pt]{article}
\def\beq{\begin{eqnarray}}
\def\eeq{\end{eqnarray}}
\def\bea{\begin{eqnarray*}}
\def\eea{\end{eqnarray*}}

\def\centeron#1#2{{\setbox0=\hbox{#1}\setbox1=\hbox{#2}\ifdim
\wd1>\wd0\kern.5\wd1\kern-.5\wd0\fi
\copy0\kern-.5\wd0\kern-.5\wd1\copy1\ifdim\wd0>\wd1
\kern.5\wd0\kern-.5\wd1\fi}}
\def\ltap{\;\centeron{\raise.35ex\hbox{$<$}}{\lower.65ex\hbox{$\sim$}}\;}
\def\gtap{\;\centeron{\raise.35ex\hbox{$>$}}{\lower.65ex\hbox{$\sim$}}\;}

\def\singleandthirdspaced{\baselineskip=\normalbaselineskip\multiply
    \baselineskip by 130\divide\baselineskip by 100}

\newcommand{\newc}{\newcommand}
\newc{\qbar}{{\overline q}}
\newc{\Kahler}{K\"ahler }
\newc{\deltaGS}{\delta_{\rm GS}}
\begin{document}
\begin{titlepage}
\begin{flushright}
{\large hep-th/yymmnnn \\ SCIPP-2008/15\\}
\end{flushright}

\vskip 1.2cm

\begin{center}

{\LARGE\bf
The Fate of Nearly Supersymmetric Vacua}

\vskip 1.4cm

{\large Michael Dine, Guido Festuccia and Alexander Morisse}
\\
\vskip 0.4cm
{\it Santa Cruz Institute for Particle Physics and
\\ Department of Physics, University of California,
     Santa Cruz CA 95064  } \\
\vskip 4pt

\vskip 1.5cm

\begin{abstract}
Supersymmetric vacua are stable.  It is interesting to ask:  how long-lived are vacua which are nearly supersymmetric?  This question is relevant if
our universe is approximately supersymmetric.  It is also of
importance for a number of issues of the physics of the landscape and eternal inflation.   In this note, we distinguish a variety of cases.
 In all of them the decay is slow.  For a flat space theory decaying to a deep AdS vacuum, the
leading behavior of the decay amplitude, if a thin
wall approximation is valid,
   is ${\cal A} = \gamma e^{-2 \pi^2/({\rm Re}~m_{3/2})^2}$ (where the phase of $m_{3/2}$ is defined in the text) for ${\rm Re}~m_{3/2}>0$, and zero otherwise.
   Metastable supersymmetry breaking generally yields parametrically more rapid decays.  For nearly supersymmetric decays, we will see that it is necessary to compute subleading terms in the exponential to extraordinarily high accuracy before one can meaningfully discuss the prefactor.
\end{abstract}

\end{center}

\vskip 1.0 cm

\end{titlepage}
\setcounter{footnote}{0} \setcounter{page}{2}
\setcounter{section}{0} \setcounter{subsection}{0}
\setcounter{subsubsection}{0}

\singleandthirdspaced

\section{The Stability of Nearly Supersymmetric Vacua}

It has long been appreciated that the state we find about us might be metastable.  This possibility has been sharpened recently by considerations of
the landscape\cite{landscape} and of metastable supersymmetry breaking\cite{iss,retrofitting}.  If our universe is unstable, we are interested in the problem of tunneling
from a state that is nearly Minkowski to a state (more precisely a big crunch) with negative cosmological constant.  The problem of the decay of flat space
to a negative c.c. state was first considered by Coleman and DeLuccia\cite{cdl}; the case of the decay of a state with a small, positive cosmological constant has been
considered by various authors\cite{parke,banks,bousso,aguirre}.

If nature were exactly supersymmetric, and space-time precisely Minkowski (i.e. zero c.c.), then, as we will review, tunneling to a lower energy
state, supersymmetric or not, would be forbidden\cite{weinberg}.   Of course, nature is not exactly supersymmetric, but it may be approximately so, with
supersymmetry broken at a scale well below some more fundamental scale (the unification scale, string scale, etc.).   The emergence of the {\it landscape}
suggests that our universe may exist in a sea of large, negative cosmological constant states:  more precisely,
with cosmological constant much larger in absolute value than $m_{3/2}^2 M_p^2$ where $m_{3/2}$ is the gravitino mass and $M_p$ the Planck mass.  The developing understanding of metastable supersymmetry
breaking suggests that there might also be some number of states with smaller, negative, cosmological constant, of order
$m_{3/2}^2 M_p^2$.

As we will explain in this paper, for the deep anti-de Sitter (AdS) states, as
$m_{3/2} \rightarrow 0$, with the cosmological constant fixed to zero,  there are various possibilities:
\begin{enumerate}
\item  The lower c.c. state is not even approximately supersymmetric, in which case the tunneling amplitude vanishes
\item  The lower c.c. state is supersymmetric or approximately so: if $m_{3/2}$ is small, and if ${\rm Re} ~m_{3/2} >0$ the state decays. If the scales
 of the potential, $M$, and changes in fields, $\Delta \phi$, are small compared to $M_p$, the bubble wall is thin,  and the tunneling amplitude is given by the universal form:
\beq
{\cal A} = \gamma ~{\rm exp}\left ( -{2 \pi^2 M_p^2 \over ({\rm Re} ~m_{3/2})^2} + \Delta \right ).
\label{universalamplitude}
\eeq
Here $\Delta$ is suppressed relative to the
leading term in the exponent by powers of $M/M_p,\Delta \phi/M_p$ and $m_{3/2}$.  We will explain
the nature of these corrections, as well as the problem of determining the prefactor $\gamma$\cite{callan}.
If Re $m_{3/2}<0$, the tunneling amplitude vanishes (if  Re $m_{3/2}=0$ then, when the tunneling amplitude
is non-zero, the exponent in the tunneling amplitude is larger and non-universal).
The phase of $m_{3/2}$ appearing in this expression requires some explanation, which we provide.
This result has been noticed in particular cases in (especially \cite{kalloshlinde}), but its generality has not been stressed.
\item  If (as we might expect to be the generic case in a landscape), changes
in fields are large,  $\Delta \phi \sim M_p$, then the thin wall approximation is not valid,
and the exponent
in the tunneling amplitude is non-universal.  In this case, the tunneling amplitude is expected to vanish, or to be suppressed by
\beq
e^{-{\beta M_p^2/m_{3/2}^2}}
\eeq
with $\beta$ a constant of order one.
\end{enumerate}

In the case of metastable supersymmetry breaking, we will see that:
\begin{enumerate}
\item  Gravitational effects in tunneling are unimportant, and tunneling is always allowed.
\item  The form of the tunneling amplitude is non-universal, depending in detail on the underlying structure.
The tunneling amplitude is typically of the form
\beq
{\cal A} \propto e^{-\left ( {M_p \over m_{3/2}}\right ) \times \left ( {M_p \over m_{3/2}} \right )^a}
\eeq
where $a$ is a fraction less than $1$.
\end{enumerate}

In the next section, we review the vanishing of tunneling in the supersymmetric limit.  This fact can be understood
from general considerations similar to those which enter in the proof of the positive energy theorem\cite{wittenpositive,deser,hull}, and can
be derived in a quite general way from semiclassical considerations\cite{weinberg} in any situation whenever
an effective field theory description is valid.   In the third section we show that for small supersymmetry
breaking, when an effective field theory description is valid, one obtains the universal expression of eqn. (\ref{universalamplitude}).
We explain that the formula has corrections of order the scales in the potential divided by $M_p$ and the
field variations (including light, hidden sector fields) divided by $M_p$, as does
the wall thickness; as the scales in the potential approach the Planck scale, the thin wall approximation (as
well as the effective action description) breaks
down. In section \ref{prefactor}, we consider the corrections to the leading,
universal result.  The nearly supersymmetric limit allows one to organize the complete determinant computation, outlined by
Callan and Coleman\cite{callan}, rather simply in the language of a three dimensional
effective field theory.  This analysis clarifies issues connected with the validity of the thin wall approximation,
and also makes clear that one cannot meaningfully discuss
the prefactor without computing the exponential in the tunneling amplitude with extraordinary accuracy.  In this section,
we comment on the case of decays with small positive cosmological constant, noting a breakdown of the effective field
theory analysis in a certain limit, which in turn implies a breakdown (known from
other work) of the thin wall approximation.  In section \ref{metastable}, we discuss
metastable supersymmetry breaking.  In the gravitational context, as we have indicated above, metastable breaking refers to situations where there
is a lower lying, supersymmetric state with cosmological constant of order $m_{3/2}^2 M_p^2$ (in absolute value).  By considering simple examples,
we will see that the decay rate, while depending on the details of the potential, is parameterically faster than the deep AdS case (though the lifetimes
can easily be extremely long).   We note that the distinction between metastable and deep AdS states fits well into known categories of dynamical
supersymmetry breaking.  In the concluding section,
 we engage in some conjectures about the significance of these results.  For the landscape, they reinforce the notion that approximately
 supersymmetric states are naturally extremely stable.  This is of interest since one might suspect that supersymmetric
 stationary points of effective actions are much less common that non-supersymmetric ones.  Stability, however, may single out such states\cite{dfkm}.
 In thinking about eternal inflation\cite{aguirreeternal,gutheternal}, there
 has been concern about the possibility of $dS$ states whose lifetime is of order the recurrence timescale.  For states which can only decay to
 deep AdS states, our results are problematic, but for what may well be the generic situation of metastable supersymmetry breaking,
 lifetimes, while long, are much shorter than recurrence times.

\section{Tunneling (or its absence) in the Supersymmetric Limit}

The absence of tunneling from flat space in the supersymmetric limit follows from the existence of global supercharges which satisfy the
standard supersymmetry algebra:
\beq
\{Q_\alpha,Q_\beta\} = P^\mu \gamma_\mu.
\eeq
Arguments based on this algebra underlie Witten's proof of the positive energy theorem\cite{wittenpositive,hull}.  But one can make a different argument\cite{weinberg},
based around the theory of semiclassical tunneling developed by Coleman\cite{coleman} and Coleman and DeLuccia\cite{cdl}.
Consider, first, the case of a general (non-supersymmetric) field theory without gravity.  Following Coleman, we suppose that the
energy difference between the false and true vacuum is $\epsilon$, where $\epsilon$ is small compared to other scales of the problem.
We are interested in the amplitude to produce a bubble.   We also take the bubble wall to be thin.  Then we can treat the size of the wall, $\rho$,
as a collective coordinate.  For large values of $\rho$, the action for this coordinate is the sum of two terms, one of which is the surface tension,
$S_1$ (in Coleman's notation) times the (Euclidean) surface area of the bubble, the second of which is equal to $-\epsilon$ times the volume of the bubble.
In other words, the action is:
\beq
B(\rho) = 2 \pi^2 S_1 \rho^3  - {1 \over 2} \pi^2 \epsilon \rho^4.
\eeq
This action has a stationary point (actually a maximum) for
\beq
\rho = 3 S_1/\epsilon.
\eeq
The presence of a negative mode gives rise to an imaginary part in the energy -- precisely the indication of an instability.  We see self-consistently
the condition for a thin wall; $\rho$ should be much larger than the other length scales of the problem (typically the Compton wave lengths of the
field involved in the transition).

\subsection{Features of Tunneling in Theories of Gravity}

In the gravitational case, we can proceed in a similar way.  Following \cite{cdl}, we can write the line element for the
$O(4)$ symmetric bounce as:
\beq
ds^2 = d\xi^2 + \rho(\xi)^2 d\Omega^2.
\label{wallmetric}
\eeq
Then, in the case of a single field, $\phi$, the equations of motion for $\phi$ and $\rho$ are:
\beq
\phi^{\prime \prime} + 3{\rho^\prime \over \rho} \phi^\prime = {d\over d\phi}U(\phi)~~~~
\rho^{\prime 2} = 1 + {1 \over 3} \kappa \rho^2 ({1 \over 2}\phi^{\prime 2} - U(\phi)).
\eeq
where $\kappa=8\pi G$.
Coleman and DeLuccia compute the action for a thin-walled bounce.  In this case, as we will discuss further, the motion of $\phi$ occurs at some characteristic,
large, value of $\rho$, $\rho = \bar \rho$, while $\rho$ changes only very slightly.
 For large $\rho$, their result may be written:
\beq
B(\bar \rho) = 2 \pi^2 (S_1 - \sqrt{4\epsilon \over 3 \kappa}) \rho^3 +{ 6 \pi^2 \rho^2 \over \kappa} + {\cal O} ( \rho).
\eeq
The fact that the energy density term grows only as $\rho^3$ is related to the well-known fact that surface areas and volumes grow similarly
in AdS space.  It is clear that $B( \rho)$ has no extremum if $\epsilon$ is too small (for fixed $S_1$).  The critical value of the ratio $\epsilon/S_1^2$
is such that
\beq
{\epsilon} = { 3 \over 4}\kappa S_1^2.
\label{criticalvalue}
\eeq

As one approaches the critical point, the radius becomes large.
Writing
\beq
\epsilon = {3 \over 4} \kappa S_1^2 (1 + \delta)
\label{deltadefinition}
\eeq
one has, for small $\delta$:
\beq
B(\rho) = - \pi^2  \rho^3 S_1 \delta + {6 \pi^2 \rho^2 \over \kappa}.
\eeq
So
\beq
\bar \rho = {4 \over \kappa S_1 \delta}~~~~B = {2 \pi^2 \bar \rho^2 \over \kappa},
\eeq
which agrees with the result of CDL in this limit.  Note that for negative $\delta$, there is no stationary point, and correspondingly no bounce.
As explained by CDL, in the case where a bubble exists, it evolves to a singular geometry.  We will not speculate on the significance of the crunch,
other than to remark that it can't be understood with low energy effective field theory.

\subsection{The Wall Thickness}

It would seem that, in the nearly critical situation $\delta \rightarrow 0_+$, the bubble becomes large and the validity
of the thin wall approximation arbitrarily good.  However, even though the wall is thin viewed in the metric of equation (\ref{wallmetric}),
the relevant measure of ``thinness" is the change in $\rho$ across the wall.  This is not surprising, since the action is sensitive
to the area of the bubble on $S_3$.

In our simple model (presented in section 3), $\Delta \rho$ is small provided all of the scales in the superpotential $M,\mu$, etc., are small compared to the Planck mass $M_p$.  More generally, the requirement
is that the potential should be small in Planck units, and similarly all field excursions ($\Delta \phi_i)$.
To see this, note:
\beq
\Delta \rho = \int_{\rho_-}^{\rho_+} d\rho = \int_{\xi_-}^{\xi_+} d\xi{d\rho \over d\xi}
\eeq
$$~~~~~\approx \int d\phi {d\xi \over d\phi} [1+{1 \over 3} \kappa \rho^2 ( \vert  \phi^{\prime 2} \vert  - U(\phi_+))]^{1/2}.$$
To get a rough estimate, can approximate the integral of $\sqrt{ \vert  \phi^{\prime 2} \vert  - U(\phi_+)}$ across the wall by
${1 \over 2} \sqrt{\epsilon} M^{-1}$, where $M^{-1}$ is roughly the wall thickness in the thin wall approximation.  Then, for large $\rho$,
we have
\beq
\Delta \rho \sim \rho {\sqrt{\epsilon} \over M M_p}
\eeq
which is of order $\rho {M \over M_p} $.  We see that if $M$ is small compared to $M_p$, the thin wall description is valid; conversely, if the scales in the
potential are comparable to $M_p$, the wall thickness is comparable to $\rho$.

\subsection{Supersymmetric and Nearly Supersymmetric Vacua}
\label{susyvacua}
In the case where both the initial and final states are supersymmetric, the bound of eqn. (\ref{criticalvalue})  is saturated.  Recall that for a supersymmetric theory, the potential
is given by
\beq
U = e^{\kappa K} \left [ D_i W \overline{D_j W} g^{ij} - 3\kappa \vert W \vert^2 \right ],
\eeq
with
\beq
D_i W = {\partial W \over \partial \phi_i} + \kappa{\partial K \over \partial \phi_i} W
\eeq
and $g^{ij}$ is the inverse of the Kahler metric, $g_{ij} = {\partial^2 K \over \partial \phi_i \phi_j^*}$ (in this section
we are choosing units with $M_p=1$).
Supersymmetry is unbroken if and only if $D_i W = 0 ~\forall~ i$.  In this case, the value of the potential
at the minimum is $-3 \kappa e^{\kappa K}\vert W \vert^2$.  Necessarily $W=0$ for the Minkowski vacuum.  We will
denote the value of the field(s) in this vacuum by $\phi_+$ (these formulas generalize immediately if there are several fields).  Calling the
value of $\phi$ in the ``other" vacuum $\phi_-$, and writing
\beq
 W(\phi_-) =\Delta W
 \eeq
 we have
$$\epsilon = 3\kappa  e^{\kappa K(\phi_-)} \vert  \Delta W \vert^2 $$
It is a well-known result that there exist static domain walls between supersymmetric Minkowski or AdS states with tension given by
\beq
S_1 = 2 \Delta e^{\kappa K/2}  W
\label{susytension}
\eeq
(we will see shortly that this statement requires refinement when the walls are not thin, in the sense used above, though it is exact
when one of the states has zero cosmological constant, i.e. vanishing $W$).
So the supersymmetric case sits precisely on the border between tunneling and no tunneling (at least for the decay of Minkowski space, or
when $M, \Delta \phi \ll M_p$).

One non-supersymmetric generalization is immediate.
In the case that the Minkowski vacuum is supersymmetric, and the AdS is not (even approximately),
it is easy to see that tunneling does not occur.  The expression of eqn. (\ref{susytension})
is, in fact, a lower (BPS) bound, which is saturated in the supersymmetric case.  If $D_i W \ne 0$ for some fields in the lower cosmological constant
state, then $\epsilon < {3\over 4} \kappa S_1^2$, so the parameter $\delta$ defined earlier in eqn. (\ref{deltadefinition}) is less than zero.

\subsection{More Careful Treatment of the BPS Bound}
\label{bps}

An elegant treatment of the BPS domain wall problem, which is directly applicable to the supersymmetric tunneling situation, is provided
in \cite{kalloshlinde}.  In particular, we can write a result for the action of the $O(4)$ symmetric bounce configurations, at large $\rho$,
which is exact, including (semiclassical) gravitational corrections.

First, consider the question of the wall thickness.  Coleman and DeLuccia ignore gravitational effects inside the wall.  We have seen that this is
a good approximation
provided $\Delta \phi/M_p \ll 1$ and the scales in the potential
are small compared to the Planck scale.  But we can consider such corrections readily in the supersymmetric case, and show that the bound of
eqn. (\ref{criticalvalue}) is still saturated.  We need, first, to give a definition of the wall in the case that the wall has finite thickness.  Again, following CDL,
call $\phi_+$ the value of the field(s) in the ``false" vacuum, and $\phi_-$ the value in the ``true" vacuum.  Define $\rho_{\pm}$ ($\xi_{\pm}$) as the value
of $\rho$ ($\xi$) at which the field $\phi$ takes a value $(1-\lambda)$ of $\phi_{\pm}$, for some small $\lambda$.  Then divide the space into three regions:
\begin{enumerate}
\item  $\rho < \rho_-$.  In this region, the bubble is essentially in the true vacuum.
\item  $\rho_- < \rho < \rho_+$.  This is the bubble wall.
\item  $\rho > \rho_+$.  In this region, the bubble is essentially in the false vacuum.
\end{enumerate}
According to CDL, the action for this system is the sum of three terms:
\begin{enumerate}
\item  From region I, we have a contribution:
\beq
S^{(1)} = - {12 \pi^2 \over \kappa} \int_0^{\rho_-} \rho d\rho [1-{1 \over 3} \kappa \rho^2 U(\phi_-)]^{1/2}
\eeq
\item  From region II, we have the tension contribution.  For large $\rho$, the problem is essentially one dimensional,
and one can take over the result for the BPS tension, derived particularly elegantly in \cite{kalloshlinde} (for earlier derivations,
see in particular\cite{domainwalltensions})
\beq
S^{(2)} = 4 \pi^2 [\rho_+^3 e^{\kappa K(\phi_+)/2} \vert W(\phi_+) \vert  - \rho_-^3 e^{\kappa K(\phi_-)/2}\vert  W(\phi_-) \vert].
\eeq
\item  It is necessary to subtract the action for the configuration with no bounce.  This gives an additional contribution:
\beq
S^{(3)} = + {12 \pi^2 \over \kappa} \int_0^{\rho_+} \rho d\rho [1-{1 \over 3} \kappa \rho^2 U(\phi_+)]^{1/2}.
\eeq
\end{enumerate}

The integrals in $S^{(1)},S^{(3)}$ are particularly simple.  Given that
\beq
U(\phi_{\pm}) =  \kappa e^{\kappa K(\phi_\pm)}\vert W(\phi_\pm)\vert^2
\eeq
for general, non-zero value of $W(\phi_\pm)$, the terms proportional to $\rho^3$ cancel in the large $\rho$ expansion of the bounce action.  If the state $\phi_+$
has vanishing cosmological constant, the cubic terms still cancel exactly; one is left with a positive quadratic contribution:
\beq
S(\rho_+) \simeq {6 \pi^2 \over \kappa}\rho_+^2
\eeq
and there is no stationary point, and hence no tunneling.
This analysis holds in any situation where an effective field theory description is valid, in particular when excursions of fields
are small compared to the Planck mass.

\section{Broken Supersymmetry in the Higher Energy State}
\label{brokensusy}

Now we consider the case where, in the higher energy, Minkowski state, supersymmetry is broken, while a much lower state with c.c. larger in absolute value than $m_{3/2}^2 M_p^2$ is approximately supersymmetric.  We will assume that there are some fields, $\phi$, with mass $M \gg m_{3/2}$; necessarily there are fields, $z$, responsible for supersymmetry
breaking, with masses less than or of order $m_{3/2}$.  We will take the superpotential to have the (hidden sector) form:
\beq
W = W_\phi(\phi) + W_z(z)
\eeq
(we will argue later that this assumption is not particularly strong), and we will assume that $\Delta z \ll M_p$.  Then we can define $W_\phi(\phi_+) = 0$,
and $W_\phi(\phi_-) = \Delta W$, and, in the Minkowski vacuum, $e^{K/2} W_z(z) \equiv e^{K/2} W_0 = m_{3/2}$.  Neglecting corrections of
order $m_{3/2} (M/M_p,\Delta \phi/M_p)$, we have that
\beq
S_1 = 2 \Delta W~~~  \Delta U = -{3 \over M_p^2} \left ( \vert \Delta W \vert^2 +2 ~ {\rm Re}~  W_0 \Delta W \right ).
\eeq
So we have
\beq
\delta = {2 {\rm Re}~(m_{3/2} \Delta W) \over \vert \Delta W \vert^2}
\eeq
and
\beq
\bar \rho = {\vert \Delta W \vert \over {\rm Re~} (m_{3/2} \Delta W)}.
\eeq
The action is
\beq
B = {2 \pi^2 \rho^2 \over \kappa}.
\eeq
Note that it is important that the light field, $z$, should not take Planckian excursions; otherwise, in addition to $W_0 \sim m_{3/2} M_p^2$, there will
be changes in the superpotential $W(\phi_-)$ of the same order.

An illustrative model which satisfies these conditions is that of \cite{dfkm}.  Here there
is a heavy field, $\phi$, and a light field, $z$.  The superpotential is taken to be simply a sum of two terms,
\beq
W(\phi,z) = {M \over 2} \phi^2 - {\lambda \over 3} \phi^3 + \mu^2 z + W_0
\label{wphiz}
\eeq
while the Kahler potential is
\beq
K = \phi^\dagger \phi + z^\dagger z - {1 \over 2 \Lambda^2} z^\dagger z z^\dagger z,
\label{kahlerwz}
\eeq
with $M \ll M_p$, $\Lambda \ll M_p$.
In the limit that $\mu = W_0 =0$, supersymmetry is unbroken.  The system has vacua at $\phi = 0$ and (at leading order in ${M\over M_p}$) $\phi = M/\lambda$.
The first has vanishing cosmological constant.  The second has
\bea
W(\phi) \equiv \Delta W = -{1 \over 6} {M^3 \over \lambda^2}.
\label{defdeltaw}
\eea
The breakup of the superpotential in eqn. \ref{wphiz} might arise, at low orders in fields, due to symmetries; the Kahler potential
of eqn. \ref{kahlerwz} for $z$ might arise if supersymmetry is broken dynamically in a hidden sector.
Now, if we turn on $\mu^2$, we see that to leading order in ${\Lambda^2 \over M_p^2}$, the vanishing of cosmological constant for the $\phi=0$ state requires that
\beq
W_0 = {1 \over \sqrt{3}} \mu^2 M_p e^{i\alpha}
\eeq
where the phase $\alpha$ is not constrained.
In this state, we have
\beq
\langle z \rangle = {\cal O} ({\Lambda^2  \over M_p}).
\eeq
and $\vert D_z W\vert^2 \sim m_{3/2}^2 M_p^2\sim \mu^4$.

If $W_0\ll \Delta W$ the second state, for which approximately $\phi={M/\lambda}$, will have a c.c. much larger in absolute value than $m_{3/2}^2 M_p^2$, and at first order in $\mu^2$:
\beq
\langle z \rangle = {\cal O} ({\mu^2 M_p^2 \over \Delta W})
\eeq
In this state we also have $\vert D_z W\vert^2 \sim \mu^4$.
As a result of our introduction of the parameter $\Lambda$, $\langle z\rangle  \ll M_p$ in both states; similarly, $\Delta \phi \ll M_p$ because $M \ll M_p$.

On the other hand, when we calculate the difference in potential between the two states, the contribution of the $3 \vert W \vert^2$ term is:
\beq
\Delta U = -3 M_p^{-2} e^{\kappa K(\phi_-)} (\vert \Delta W \vert^2 + 2 {\rm Re}~\Delta W W_0) + {\cal O} \left({W_0\over M_p^2}\right)^2.
\eeq
where $\Delta W$ is given by eqn. (\ref{defdeltaw}).
As we have seen, near the supersymmetric limit the thin wall approximation is valid provided that $\vert \Delta z \vert/M_p,\;\vert \Delta\phi \vert/M_p \ll 1$.  So from the energy difference calculated above,
and the knowledge of the minimum tension domain wall, we can obtain the tunneling amplitude.  In this model,
the tension of the domain wall separating the two vacua remains $S_1 = 2 e^{\kappa K(\phi_-)} \Delta W$, up to terms of order $m_{3/2}^2$ and $\Delta z/M_p$, $\Delta \phi/M_p$.
To see this, note that:
\begin{enumerate}
\item  Because $\phi,z \ll M_p$, we can ignore the $e^{\kappa K}$ factors multiplying $W_0$ (not $\Delta W$).
\item  The $\vert DW \vert^2$ terms are of order $\mu^4 \sim m_{3/2}^2 M_p^2$
\item     The shifted $-3 \vert W \vert^2$ terms do include potential contributions of
order $m_{3/2}$.  But in fact they make no additional contribution to the tension.  This can be understood by noting that the problem at hand is, to order $m_{3/2}$,
just a deformation of the original problem in which a constant of order $m_{3/2} M_p^2$ has been added to the superpotential.  From the expressions of
\cite{kalloshlinde}, as we have discussed, the tension remains $2 e^{\kappa K(\phi_-)}\Delta W$, up to terms of order $\Delta \phi/M_p$.
\item One might also worry about the kinetic terms for $z$, but
the resulting contributions
to the tension are of order $m_{3/2}^2$.   This is easily understood in the particle analogy.  The, field $\phi$ moves from
the neighborhood of the true vacuum to the false vacuum in a time of order $M^{-1}$.  So as this field is approaching its endpoint, the light field
is just beginning its motion.  It has just enough kinetic energy to reach the ``top of the hill" in the inverted potential; the action for this
part of the motion is simply $m_Z^2 \Delta z^2$, which is suppressed in our model, in which the change in $z$ is small.
\end{enumerate}

So the leading contribution to $B(\rho)$ for large $\rho$ is obtained by using the leading order tension, $S_1 = 2 \Delta e^{K(\phi)} W(\phi)$, and
the energy splitting as described above, and
\beq
\bar \rho = {\rm Re}~ m_{3/2}~~~ B = {2 \pi^2} {M_p^2 \over ({\rm Re}~ m_{3/2})^2} \left (1 + {\cal O}(M/M_p,m_{3/2}/M_p) \right ).
\eeq

While derived in a specific model, this result is quite general.  The important ingredients is  the existence of a set of heavy fields (typical mass,
$M$), whose potential has multiple stationary points, and a small scale of supersymmetry breaking, $M \gg m_{3/2}$; inevitably this requires
some set of fields like $z$ light compared to $M$.

To summarize, we have established that:
\begin{enumerate}
\item  For an approximately supersymmetric, Minkowski vacuum, decaying to an approximately supersymmetric, AdS vacuum, the tunneling amplitude
behaves as $e^{-{2 \pi^2 A  {M_p^2 \over \vert m_{3/2}\vert^2}}}$, with $A$ an order one constant or vanishes.  For $\Delta \phi \ll M_p$ for all fields
(including hidden sector fields), and for scales in the potential $M \ll M_p$,
\beq
{\cal A} = e^{-2 \pi^2  {M_p^2 \over ({\rm Re}~ m_{3/2})^2}}.
\eeq
for positive $m_{3/2}$, and vanishes for negative ${\rm Re}~m_{3/2}$.
\item  For an approximately supersymmetric, Minkowski vacuum decaying to a non-supersymmetric
AdS state, the decay amplitude vanishes (for small $m_{3/2}$).
\end{enumerate}

\section{Small Fluctuations About the Bounce Solution}

In order to understand the prefactor in the  tunneling amplitude, it is necessary to understand the determinant of small fluctuations.  As we will
see, the question of the prefactor is only interesting if one can compute the exponent with extraordinary accuracy, including both classical and quantum
parts.  But consideration of the fluctuations is useful for understanding the thin wall (or other) approximation(s) and assessing its (their)
validity.  The case without gravity
is particularly simple, at least conceptually; the introduction of gravity raises new issues.

\subsection{The Prefactor and the Exponent Without Gravity}
\label{prefactor}

Callan and Coleman\cite{callan}, in a classic paper, outlined the problem of computing the functional determinant in the case of vacuum decay.
One might think that the decay of nearly supersymmetric states would provide a particularly simple framework in which to do such a computation.
As we now explain, however, the problem of isolating the dimensionful prefactor
is a hard one, precisely because, in this limit, the tunneling amplitude is so small.

The issues can be understood by considering
tunneling in the absence of gravity.
We focus on a theory in which there is a massive field, with mass $M$, (e.g. the massive adjoint of the simplest $SU(5)$ grand unified theory) and a hidden sector
responsible for supersymmetry breaking, with characteristic scale $\mu$.  In the limit that the supersymmetry breaking vanishes, there often are BPS domain walls,
with tension of order $M^3$.  The domain wall possesses
a bosonic zero mode arising from translations of the domain wall; in the case of supersymmetric, BPS domain walls,
there is also a fermionic zero mode.  These zero modes are described by a $2 +1$ dimensional field theory living on the wall. The effective
field theory possesses two real zero modes.  These zero modes have small components of the hidden sector fields; in general, the light fields
in this sector are decoupled from the modes on the wall.

Anticipating that the radius of the bubble is large, we can describe the light modes of the system by a $2+1$ dimensional field theory
in a background metric.  The tension is the cosmological constant of this theory; there is a contribution to the potential arising from
the bulk term in the energy.  Described in this way, the condition on the bubble radius is the condition for vanishing tadpole
in the field theory.

It is easy to write down complete sets of eigenfunctions of the three dimensional laplacian and (massless) Dirac operator\cite{callan}.  The low lying
modes include a negative mode -- up to a normalization factor, just the second derivative of the action with respect to $\rho$.
There are four translation zero modes, and then a tower of higher modes.  If one examines the
the resulting determinant, one immediately encounters a linear divergence.  This divergence is readily understood by considering
the structure of the effective action.  In the spherical geometry under
consideration, the effective action can contain a term
\beq
\int d^3 x \sqrt{g} R
\eeq
where $g_{\mu \nu}$ is the metric of the sphere, and $R$ the curvature scalar.  Such a term is not forbidden by supersymmetry, and will appear
at one loop (without supersymmetry, there is a cubic divergence, associated with an infinite renormalization of the tension).  This indicates
that the determinant depends on the microscopic details, and is neither universal or simple.

Considering, more generally, the terms which can appear in the three dimensional
effective action, one sees immediately that there are several terms in the action
which one must compute accurately before one can meaningfully discuss the prefactor.  One expects, even
classically, corrections to the leading $1/\epsilon^3$ terms
in the action behaving as $1/\epsilon^2,1/\epsilon$.
These exponentially important corrections totally overwhelm any modest effects in the prefactor associated
with the few lowest modes.
The low eigenvalues on the sphere are of order $1/\rho^2$, which for
the flat space problem, is of order $\epsilon^2.$

\subsection{Small Fluctuations in Theories Including Gravity}

Theories with gravity raise new conceptual and technical issues.  One might hope that
the nearly supersymmetric limit we have studied here is in some ways simple.  As $\delta \rightarrow 0$, the bounce radius
becomes large.   There is a relatively clean separation of light ($1/\rho \sim \delta$) and heavy ($M$) modes,
and again other light modes (with masses ${\cal O}(m_{3/2})$, from the hidden
sector) are very
weakly coupled to the modes on the wall.  Focussing,
first, on ${\rm Minkowski}~\rightarrow ~{\rm AdS}$ decays, the lightest, negative eigenvalue is of order $m_{3/2}$ in this limit.
So there does not appear to be anything singular as one approaches this point.

As in the non-gravitational case, sensibly defining the prefactor requires, first, a high degree
of precision in the calculation of the exponent.  As in that case, there is a linear divergence in the
determinant, associated with the generation of an $\int \sqrt{g}~ R$ term in the effective theory.
Such a term corresponds to an
 or $M_p/m_{3/2}$ term in the action, a small fractional correction to the leading term. but a huge
 correction to the overall rate.  More generally, writing the exponential as:
\beq
B = \alpha(M_p^2/m_{3/2}^2) + \beta (M_p /m_{3/2}) + C,
\eeq
$\alpha$ and $\beta$, for example (which include corrections in powers of $M/M_p$), must be known to better than parts in $(m_{3/2}/M_p)^2,(m_{3/2}/M_p)$ accuracy,
respectively, and
$C$ must be calculated to better than order one accuracy before a prefactor such as $\gamma m_{3/2}^4$ can be meaningfully
presented with a stated uncertainty,  $\Delta \gamma$.
Whether this is possible, even in principle (e.g. worrying about questions of convergence of the various perturbation expansions in play
here) is an interesting, if academic, question.

A potential breakdown of the thin wall approximation, signalled by difficulties with the effective action,
is provided by
the case of
of $dS \rightarrow ~{\rm Minkowski}$ decays studied in CDL.  Here, CDL find for the bounce action
\beq
B(\rho) =2 \pi^2 S_1\rho^3 + {12 \pi^2 \over \kappa} \left [{1\over \kappa \epsilon} -{1 \over \kappa \epsilon}(1-{1 \over 3} \kappa \rho^2 \epsilon)^{3/2} +{\rho^2 \over 2} \right ]
\label{dsaction}
\eeq
with stationary point at:
\beq
\bar \rho = {12 S_1 \over 4 \epsilon + 3 \kappa S_1^2}
\eeq
and
\beq
B ={216 \pi^2 S_1^4 \over \epsilon [4 \epsilon + 3 \kappa S_1^2]^2}.
\eeq

This result has puzzling features.  In particular, if we hold $S_1$ fixed and vary $\epsilon$, we see that
there is a critical value of $\epsilon$ where  the stationary point goes from being a maximum of the potential (yielding the
negative mode in the functional integral) to a minimum.  Examining eqn. \ref{dsaction},  we see that at this point, the thin wall approximation is breaking down; the second derivative
of $B$ diverges, and the separation of light and heavy modes is no longer valid.  Beyond this point, even classically, it is necessary to study the behavior
of the full equations in order to establish the existence of solutions.  In \cite{banks,bousso}, for example, it is shown numerically that solutions exist quite generally, not only
for the decays to Minkowski space, but for decays to AdS, in cases with small cosmological constant and $\delta < 0$, with decay amplitude bounded by
the exponential of the dS entropy.

\section{Metastable Supersymmetry Breaking}
\label{metastable}

In the past few years, beginning with the work of \cite{iss}, there has been a growing appreciation that metastable supersymmetry breaking is generic.  In the landscape context,
where inevitably the low cosmological constant state in which we find ourselves is surrounded by states of large, negative cosmological constant, this statement
requires some refinement.  The issues can be illustrated by a simple model of a single field:
\beq
W = \mu^2 z + {1 \over (n+3)} M^{n} z^{n+3} + W_0~~~~ K = z^* z - {1 \over 2\Lambda^2} (z^* z)^2~~~\Lambda \ll M, M_p.
\label{metastablew}
\eeq
Here $M$ is some large energy scale; it might be as large as $M_p$, but could be much smaller.
This model has a non-supersymmetric ground state near the origin, as we have discussed above; choosing $\vert W_0\vert = {1 \over \sqrt{3}} \mu^2$ renders the
cosmological constant zero in this state.  But the system also has a supersymmetric ground state, with
\beq
z^{n+2} = \mu^2 M^n.
\eeq
This state has a negative cosmological constant,
\beq
V_0 = -3 e^{K}{ \vert W_0 \vert^2 \over M_p^2}  (1+ {\cal O} (({\mu \over  M_p})^{2/3})
\eeq
In other words, the state is lower than the non-supersymmetric, zero cosmological constant state by an amount  $3 m_{3/2}^2 M_p^2$.

For potentials such as this, as we will see, gravitational corrections are not important (even for $M \sim M_p$), but a thin wall analysis
is not appropriate.  One can estimate the bounce action, instead, by taking derivatives of fields to be:
\beq
{d \phi \over dr} \sim {\Delta \phi \over R}
\eeq
where $R$ is the bubble size, which we assume comparable to the thickness.  Calling the difference in energy of the states $\Delta E$, this gives
\beq
R \sim {\Delta \phi \over \sqrt{ \Delta E}}
\eeq
and
\beq
S_b \sim {(\Delta \phi)^4 \over \Delta E}.
\eeq
This is much larger than $(M_p^2/m_{3/2})^2$ provided $\Delta \phi \ll M_p$.
Note that the Hubble constant in the AdS state is of order $m_{3/2}$, so as long as $\Delta \phi \ll M_p$,
gravitational corrections are unimportant.
For our model above, this gives
\beq
S_b \approx \left ( {M \over \mu} \right )^4 \left [{\mu\over M} \right ]^{8 \over n+2} .
\eeq
We have written the amplitude in this form to indicate that even for $M \sim M_p$, tunneling is more rapid than in the
deep AdS case we have discussed above.

\subsection{Models With Stable and Metastable Dynamical Supersymmetry Breaking}

We expect that the model of eqn. (\ref{metastablew}) captures the principal features of most models of metastable dynamical supersymmetry breaking:
gravitational effects are unimportant, and tunneling is parameterically faster than in the deep AdS case.
It certainly fits in the class of retrofitted models\cite{retrofitting}.  The ISS model has a similar structure\cite{iss}.  In that case, the small
parameter is the quark mass, $m$, over the dynamical scale $\Lambda$, and the tunneling amplitude behaves as a power of $m/\Lambda$.

What about models such as the "3-2" model, with what might be called "stable supersymmetry breaking"?  Even in this model, including
non-renormalizable interactions, one expects that there will be supersymmetric vacua.  The critical distinction lies in the fact that in this case, the
minima of the potential lie at fields of order $M$, where $M$ is the
scale of the non-renormalizable operators; the depth of the corresponding AdS states then does not tend to zero as $m_{3/2} \rightarrow 0$.
This behavior seems generic, but there are curious exceptions.  In particular, consider the model with gauge group $SU(5)$ and
a single $\bar 5$ and $10$.  In this model, there is simply no invariant superpotential one can write down\cite{adssu5}.  So there is no
lower energy supersymmetric state.  Of course, embedded in some larger landscape, other behaviors are possible.

\section{Implications}
\label{conclusions}

We have seen that, in the nearly supersymmetric limit, it is possible to make general statements about vacuum tunneling.
In the limit that field excursions are small compared to the Planck scale, and the scales of the potential are small,
we have seen that there is a simple, universal form of the tunneling amplitude.  More generally, we have understood
the parametric suppression of tunneling both in the case of what we have called ``deep AdS" vacua, and vacua
with what we have defined to be metastable supersymmetry breaking.

We have also seen that in this limit (and also in global supersymmetry, in nearly supersymmetric systems), the problem
of tunneling can be usefully phrased in the language of three dimensional field theory.  This provides a useful
setup in which to consider the calculation of the functional determinant, and to assess the validity of the thin wall
approximation.  In the gravitational case, further study of the effective field theory is warranted (e.g. not all of the low
lying modes are actually localized on the bubble wall) and will be reported elsewhere.

If a landscape picture of fundamental physics is valid, the observations here might be of some importance.  They reinforce the notion
that supersymmetric states in a landscape are special, in that they are automatically highly metastable.   Few, if any, other generic
features of stationary points of known string theories lead to stability in such a simple way\cite{dfkm}.

Our results might also be relevant for considerations of eternal inflation in a landscape.  It is natural to ask, for example, are lifetimes
of typical states long or short compared to recurrence times\cite{frievogel}.  If the states with metastable supersymmetry breaking dominate
the landscape, than this is not an issue.  If the deep AdS states are typical, then those for which tunneling vanishes for vanishing c.c.
will have lifetimes of order the recurrence time.

\clearpage

\noindent
{\Large  {\bf Acknowledgements}}

We acknowledge valuable conversations with Steve Shenker; it was questions from Steve which stimulated this work.  We
also are appreciative of conversations and advice from Anthony Aguirre and Ben Frievogel.  Tom Banks and Matt Johnson
offered comments on an early version of this manuscript which lead us to refine a number of the ideas presented
here.    This work supported in part by the U.S.
Department of Energy.


\begin{thebibliography}{99}

\bibitem{landscape}  The notion of a landscape owes much of its force to the work of Weinberg on the cosmological constant,
  S.~Weinberg,
  Rev.\ Mod.\ Phys.\  {\bf 61}, 1 (1989); the first plausible realization of Weinberg's scenario in turns of flux vacua was due to
  Bousso and Polchinski;
  R.~Bousso and J.~Polchinski,
  JHEP {\bf 0006}, 006 (2000)
  [arXiv:hep-th/0004134];
a more complete picture realizing the flux landscape is that of KKLT:
  S.~Kachru, R.~Kallosh, A.~Linde and S.~P.~Trivedi,
  Phys.\ Rev.\  D {\bf 68}, 046005 (2003)
  [arXiv:hep-th/0301240].
Susskind coined the term,
  L.~Susskind,
  arXiv:hep-th/0302219.
and provides a conceptual overview in his book
L.~Susskind, {\it The Cosmic Landscape}, Little Brown, New York, 2005.



\bibitem{iss}
  K.~A.~Intriligator, N.~Seiberg and D.~Shih,
  JHEP {\bf 0604}, 021 (2006)
  [arXiv:hep-th/0602239].



\bibitem{retrofitting}
  M.~Dine, J.~L.~Feng and E.~Silverstein,
  Phys.\ Rev.\  D {\bf 74}, 095012 (2006)
  [arXiv:hep-th/0608159].

\bibitem{cdl}
  S.~R.~Coleman and F.~De Luccia,
  Phys.\ Rev.\ D {\bf 21}, 3305 (1980).





\bibitem{parke}
  S.~J.~Parke,
  Phys.\ Lett.\  B {\bf 121}, 313 (1983).



\bibitem{banks}
  A.~Aguirre, T.~Banks and M.~Johnson,
  JHEP {\bf 0608}, 065 (2006)
  [arXiv:hep-th/0603107].

\bibitem{bousso}
  R.~Bousso, B.~Freivogel and M.~Lippert,
  Phys.\ Rev.\  D {\bf 74}, 046008 (2006)
  [arXiv:hep-th/0603105].


\bibitem{aguirre}
  A.~Aguirre and M.~C.~Johnson,
  Phys.\ Rev.\  D {\bf 77}, 123536 (2008)
  [arXiv:0712.3038 [hep-th]].


\bibitem{weinberg}
  S.~Weinberg,
  Phys.\ Rev.\ Lett.\  {\bf 48}, 1776 (1982).



\bibitem{callan}
  C.~G.~.~Callan and S.~R.~Coleman,
  Phys.\ Rev.\  D {\bf 16}, 1762 (1977).


\bibitem{kalloshlinde}
  A.~Ceresole, G.~Dall'Agata, A.~Giryavets, R.~Kallosh and A.~Linde,
  Phys.\ Rev.\  D {\bf 74}, 086010 (2006)
  [arXiv:hep-th/0605266].









\bibitem{wittenpositive}
  E.~Witten,
  Commun.\ Math.\ Phys.\  {\bf 80}, 381 (1981).



\bibitem{deser}
  S.~Deser and C.~Teitelboim,
  Phys.\ Rev.\ Lett.\  {\bf 39}, 249 (1977).


\bibitem{hull}
  C.~M.~Hull,
  Commun.\ Math.\ Phys.\  {\bf 90}, 545 (1983).


\bibitem{dfkm}
  M.~Dine, G.~Festuccia, A.~Morisse and K.~van den Broek,
  JHEP {\bf 0806}, 014 (2008)
  [arXiv:0712.1397 [hep-th]].



\bibitem{aguirreeternal}
  A.~Aguirre,
  arXiv:0712.0571 [hep-th].

\bibitem{gutheternal}
  A.~H.~Guth,
  J.\ Phys.\ A  {\bf 40}, 6811 (2007)
  [arXiv:hep-th/0702178].


\bibitem{coleman}
  S.~R.~Coleman,
  Phys.\ Rev.\  D {\bf 15}, 2929 (1977)
  [Erratum-ibid.\  D {\bf 16}, 1248 (1977)].




\bibitem{domainwalltensions}
  M.~Cvetic, S.~Griffies and S.~J.~Rey,
  Nucl.\ Phys.\  B {\bf 381}, 301 (1992)
  [arXiv:hep-th/9201007].





\bibitem{adssu5}
  I.~Affleck, M.~Dine and N.~Seiberg,
  Phys.\ Lett.\  B {\bf 137}, 187 (1984).


\bibitem{frievogel}
  B.~Freivogel and M.~Lippert,
  arXiv:0807.1104 [hep-th].


\end{thebibliography}
\end{document}